\documentclass{ws-mpla}

\begin{document}

\markboth{K. Nagata}
{Structure of the nucleon and Roper Resonance with Diquark Correlations}

\def\Slash#1{/\hspace{-0.23cm}{#1}} 
\newcommand{\beq}{\begin{eqnarray}}
\newcommand{\eeq}{\end{eqnarray}}
\newcommand{\bra}{\langle}
\newcommand{\del}{\partial}
\def\gsim{\displaystyle\mathop{>}_{\sim}}
\def\lsim{\displaystyle\mathop{<}_{\sim}}

\def\beq{\begin{eqnarray}}
\def\eeq{\end{eqnarray}}
\def\nn{\nonumber}
\def\pif{(4\pi)^2}
\def\SU{\mbox{SU}}
\def\scdot{\! \cdot \!}
\def\scdots{\! \cdots \!}
\def\si{\sin\theta}
\def\co{\cos\theta}
\def\ln{\mbox{ln}}
\def\n{\\ \indent}
\def\Tr{\mbox{Tr}}

\def\befc{\begin{figure}[h]\begin{center}}
\def\eefc{\end{center}\end{figure}}

\def\up{\uparrow} \def\down{\downarrow}

\def\bra{\langle}    \def\ket{\rangle}  

\def\intx{\int_0^1dx} \def\inty{\int_0^1dy} \def\intz{\int_0^1dz}
\def\loopint{\int\frac{d^4k}{(2\pi)^4}}
\def\feynint2{\int_0^12xdx\int_0^1dy}

\def\V{\mbox{V}}     \def\A{\mbox{A}}

\def\rarrow{\rightarrow} \def\larrow{\leftarrow}
\def\taup{\vec{\tau}}
\def\vecv{\vec{v}}
\def\veca{\vec{a}}
\def\gv{g_v} \def\ga{g_A}
\def\mn{M_N}         
\def\ms{M_s}
\def\ma{M_A}
\def\mq{m_q}
\def\eqs{&\hspace{-3mm}}
\def\pr{{\prime}}
\def\tr{\,\hbox{tr}\,}   
\def\calA{{\cal A}}    \def\calN{{\cal N}}
\def\calB{{\cal B}}    \def\calO{{\cal O}}
\def\calC{{\cal C}}    \def\calP{{\cal P}}
\def\calD{{\cal D}}    \def\calQ{{\cal Q}}
\def\calE{{\cal E}}    \def\calR{{\cal R}}
\def\calF{{\cal F}}    \def\calS{{\cal S}}
\def\calG{{\cal G}}    \def\calT{{\cal T}}
\def\calH{{\cal H}}    \def\calU{{\cal U}}
\def\calI{{\cal I}}    \def\calV{{\cal V}}
\def\calJ{{\cal J}}    \def\calW{{\cal W}}
\def\calK{{\cal K}}    \def\calX{{\cal X}}
\def\calL{{\cal L}}    \def\calY{{\cal Y}}
\def\calM{{\cal M}}    \def\calZ{{\cal Z}}

\def\tilF{\tilde F}
\def\bra{\langle}    \def\ket{\rangle}
\def\del{\partial}
\def\ms{M_S} \def\mq{m_q} \def\ma{M_A}
\def\ora{\overrightarrow} \def\ola{\overleftarrow}
\def\half{\frac{1}{2}}
\def\taubm{{\bm \tau}}
\def\ds{\displaystyle}
\def\eqs{&\hspace{-3mm}}
\def\intx{\int_0^1dx} \def\inty{\int_0^1dy} \def\intz{\int_0^1dz}
\def\intk{\int\frac{d^4k}{(2\pi)^4}}

\def\suf{SU_{SF}(4)}
\def\sux{SU_{SF}(6)}

\def\mltp#1#2{{\bf #1}_{\mbox{#2}}}
\def\inner#1#2{{\bm #1}\cdot {\bm #2}}
\def\outer#1#2{{\bm #1}\times {\bm #2}}

\def\diag{\mbox{diag}}
\catchline{}{}{}{}{}

\title{STRUCTURE OF THE NUCLEON AND ROPER RESONANCE WITH DIQUARK CORRELATIONS}


\author{K. NAGATA}

\address{Department of Physics, Chung-Yuan Christian University, Chung-Li 320, Taiwan\\
nagata@phys.cycu.edu.tw}

\author{A. HOSAKA}

\address{Research Center for Nuclear Physics,  Osaka University,
 Ibaraki 567-0047, Japan.
}

\maketitle


\begin{abstract}
We investigate the electric form factors of the nucleon and Roper 
resonance using a quark-diquark model.
We find that the charge radii of the nucleon and Roper resonance 
are almost the same in size.

\end{abstract}

\ccode{PACS Nos.:11.30.Rd; 13.40.Gp; 14.20.Gk. }

\section{Introduction}

Diquarks have been attracted attention in hadron physics, which play important 
roles in several phenomena, such as the ratio of 
the structure functions $\ds{\lim_{x\to 1}} F_2^n(x)/F_2^p(x)\to 1/4$,  $\Delta I=1/2$ 
rule in semi-leptonic weak decays, recent exotic studies and so 
on~\cite{Anselmino:1992vg}. 

We have proposed a description of the Roper resonance $N(1440) (P_{11})$
as a partner of the nucleon~\cite{Nagata:2005qb,Nagata:2007mk}.
In diquark models, the nucleon is described in two types of quark-diquark channels
corresponding to two kinds of the diquarks; the scalar diquark $(I(J)^P=0(0)^+)$
and axial-vector diquarks $(1(1)^-)$. Assuming the mass difference between the 
scalar and axial-vector diquarks to be caused by the spin-spin interaction 
between quarks, the two types of the quark-diquark channels have the mass difference of the order 
of the spin-spin interaction.
It is known that the mass difference between the two diquarks is also 
a main source of the $N-\Delta$ mass difference of an order of 300 MeV. 
Hence the two quark-diquark bound states have this amount of the mass difference.
Note that when a system would be exactly spin-flavor symmetric, one of the two nucleons 
is forbidden by Pauli-principle and the nucleon and $\Delta$ resonance 
would degenerate because they belong to the same spin-flavor multiplet.
In the present paper, we report recent results on the electric structure of the nucleon
and Roper resonance in the quark-diquark model~\cite{Nagata:2007mk}.

\section{Framework}

We recapitulate the chiral quark diquark model, for
details see Ref.~\cite{Nagata:2005qb,Nagata:2007mk,AbuRaddad:2002pw}
We start from the SU(2)$_R\times$ SU(2)$_L$ chiral quark-diquark model 
defined by, 
\begin{eqnarray}
{\cal L} = \bar{\chi}_c(i\rlap/\del - m_q) \chi_c +D^\dag_c (\del^2 + M_S^2)D_c
 +
{\vec{D}^{\dag \mu}}_c 
\left[  (\del^2 + M_A^2)g_{\mu \nu} - \del_\mu \del_\nu\right]
\vec{D}^{\nu}_c +{\cal L}_{I},
\label{lsemibos}
\end{eqnarray}
where $\chi_c$, $D_c$ and $\vec{D}_{\mu c}$ are the constituent quark, scalar
diquark and axial-vector diquark fields,  $m_q$, $M_S$ and $M_A$
are their masses. The indices $c$ represent the color. Note that the 
diquarks microscopically correspond to the quark bi-linears: $D_c\sim
\epsilon_{abc}\tilde{\chi}_b\chi_c,\ \vec{D}_{\mu c}\sim \epsilon_{abc}\tilde{\chi}_b 
\gamma_\mu \gamma_5 \vec{\tau}\chi_c$, where 
$\tilde{\chi}=\chi^T C\gamma_5 i\tau_2$.  
Both the diquarks belong to color anti-triplets and baryons to singlets.
The term ${\cal L}_{I}$ is the quark-diquark interaction, which is written as
\begin{eqnarray}
{\cal L}_{I}&=&G_S\bar{\chi}_cD^\dagger_c
D_{c^\prime}\chi_{c^\prime}+v(\bar{\chi}_cD^\dagger_c\gamma^\mu\gamma^5
\vec{\tau}\cdot\vec{D}_{\mu c^\prime} \chi_{c^\prime}+\bar{\chi}_c\gamma^\mu\gamma^5
\vec{\tau}\cdot\vec{D}^\dagger_{\mu c}
D_{c^\prime}\chi_{c^\prime})\nonumber \\
&&+G_A\bar{\chi}_c\gamma^\mu\gamma^5
\vec{\tau}\cdot\vec{D}^\dagger_{\mu c}
\vec{\tau}\cdot\vec{D}_{\nu c^\prime}\gamma^\nu\gamma^5 \chi_{c^\prime},
\label{eq:twoc}
\end{eqnarray}
where $G_S$ and $G_A$ are the coupling constants for the quark and scalar
diquark, and for the quark and axial-vector diquark, which
describe two kinds of bare nucleons as quark-diquark bound states: a bound state of a quark and a scalar-diquark, 
and of a quark and an axial-vector diquark.  While $v$ causes the mixing between the scalar and axial-vector channels.
We showed in Ref.~\cite{Nagata:2005qb} that the masses of the nucleon and 
Roper resonance are reproduced with suitable strengths of the interactions
$G_S$, $G_A$ ans $v$. We also showed that a mixing angle of the scalar
and axial-vector channels in the baryons is small; the nucleon (Roper) is 
a scalar (axial-vector) diquark dominant state, which is crucial in 
the structures of the nucleon and Roper resonance.

The electromagnetic interactions of the nucleon and Roper resonance are introduced in 
the Lagrangian Eq.~(\ref{lsemibos}) through the gauge couplings of 
the quark and diquarks~\cite{Nagata:2007mk}.
We effectively include the intrinsic sizes of the diquarks: a monopole type 
for the scalar diquark and dipole for the axial-vector diquark.
The size parameters of the form factors are determined so as to reproduce the 
charge radii of the proton and neutron.
In this model setup, we discuss the electric structures of the nucleon and 
Roper resonance.

\section{Results}

\begin{figure}[hbt]
\centering{
\includegraphics[width=5cm]{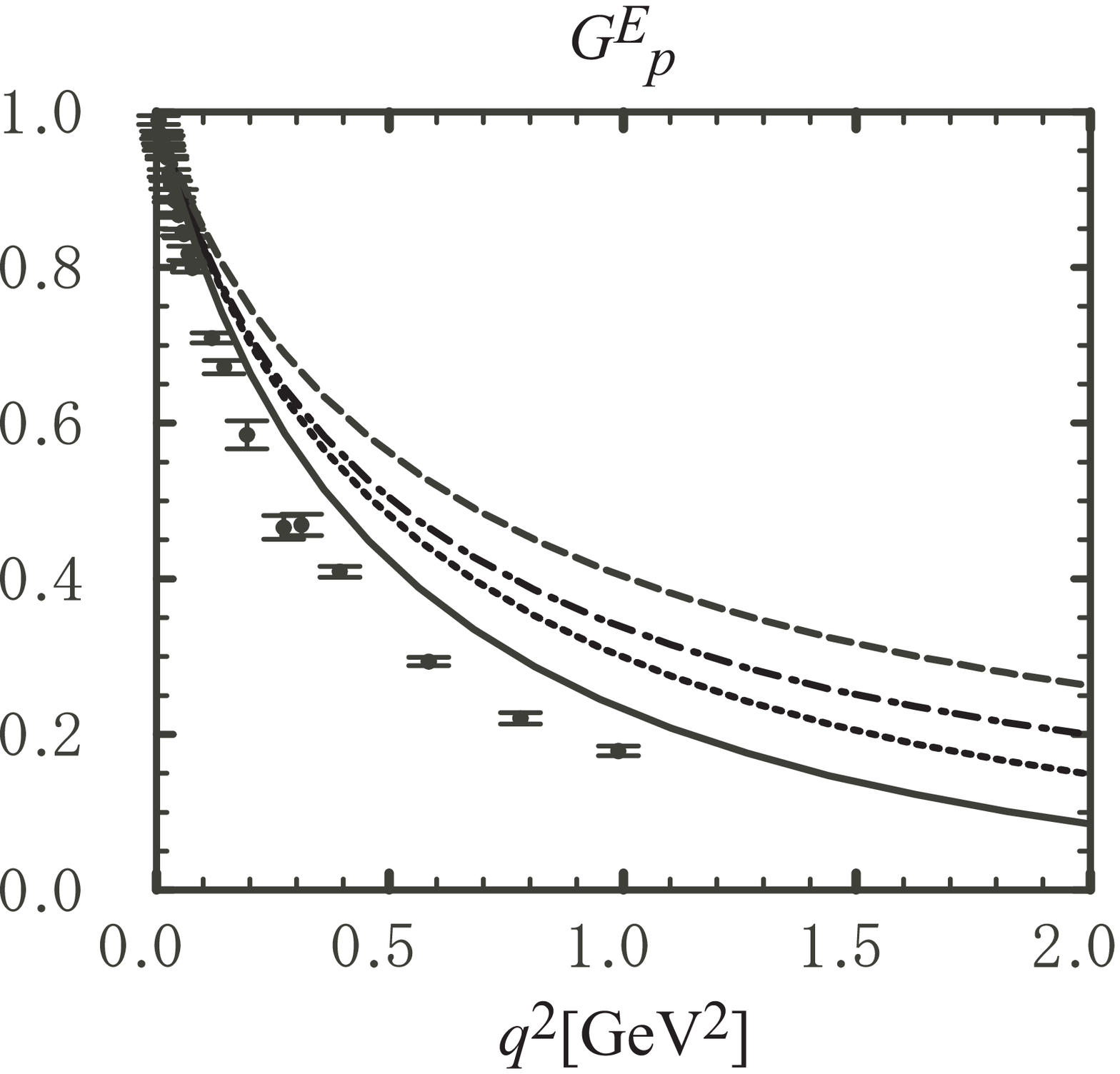}
\includegraphics[width=5.5cm]{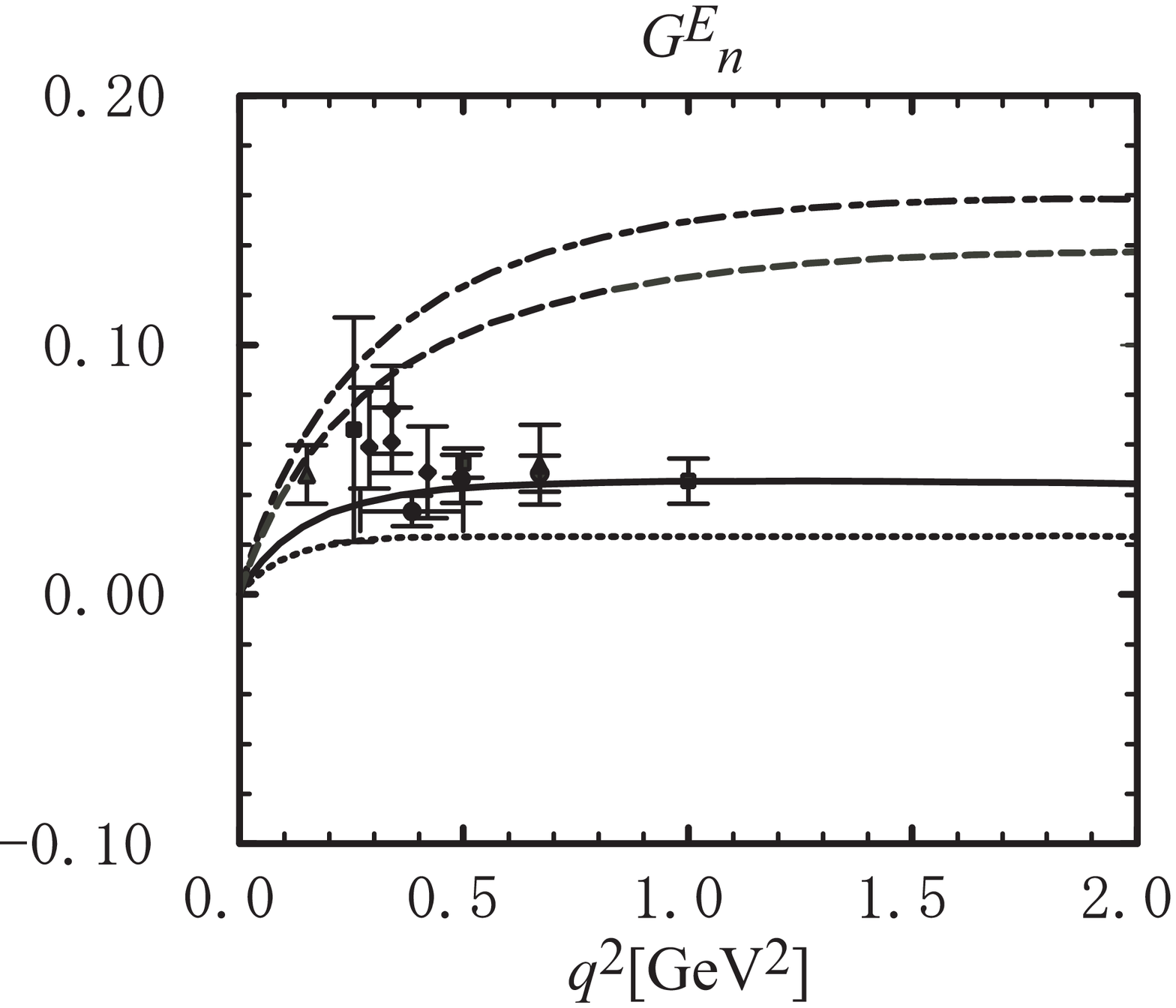}}
\begin{minipage}{11cm}
\caption{\small The electric form factors of the proton (left) and neutron
(right). The dotted, dotted-dashed, solid and dashed lines are for the cases (i), (ii), (iii) and (iv).  
The experimental data are taken 
from~\protect\cite{Hohler:1976ax,Eden:1994ji}.}\label{fig:idff}%
\end{minipage}%
\end{figure}%
Figures~\ref{fig:idff} show the electric form factors of the proton and neutron. 
The four lines show the effects of the intrinsic diquark form factors; 
case (i) includes the intrinsic form factor only for the scalar diquark, case (ii) only 
for the axial-vector diquark, case (iii) for both the diquarks and case (iv) for
neither the scalar nor axial-vector diquarks. 

The inclusion of the scalar diquark form factor enhances the slope of $G^E_p$ at $q^2=0$ , 
and suppresses that of $G_n^E$, which implies that the charge radii both of the proton and 
neutron become larger. 
With respect to the axial-vector diquark, both the slopes of $G^E_p$ and $G^E_n$ 
become larger, which implies that the charge radius of the proton becomes large while 
that of the neutron becomes smaller.
These behaviors are understood as follows. 
The scalar diquark ($ud$) carries the positive charge $+1/3$ both in the proton 
and neutron, while the axial-vector diquark contains three components 
with the electric charge $(uu,ud,dd)=(+4/3,1/3,-2/3)$.  In average the axial-vector 
diquark in the proton (neutron) carries a positive (negative) charge.
Including both the scalar and axial-vector diquark 
form factors, we obtain the reasonable results both for $G^E_p$ and $G^E_n$ 
in the region $0\leq \vec{q}^{\;2} \leq 1$ [GeV$^2$].

\begin{figure}[hbt]
\centering{
\includegraphics[width=5cm]{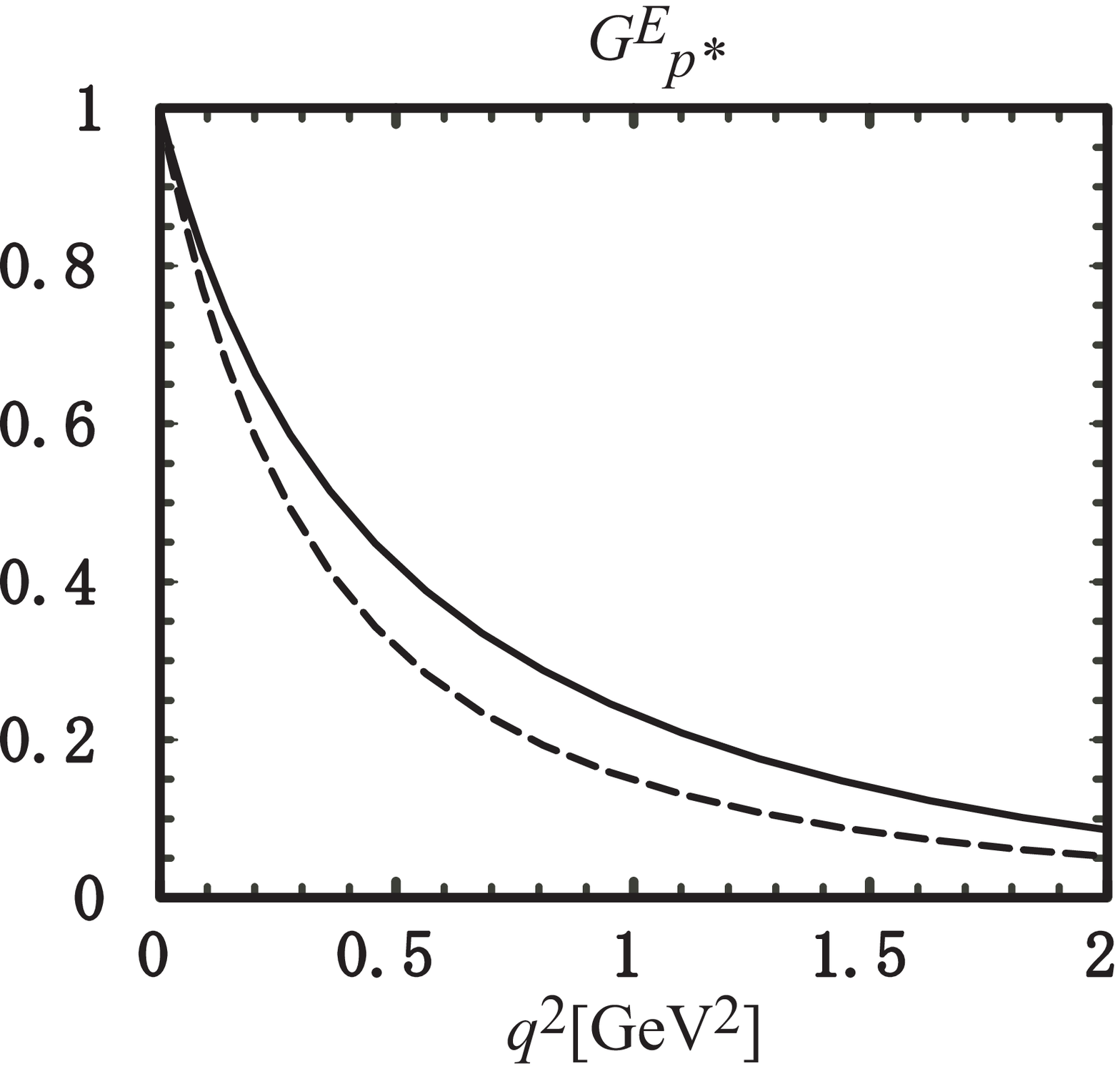}
\includegraphics[width=5.15cm]{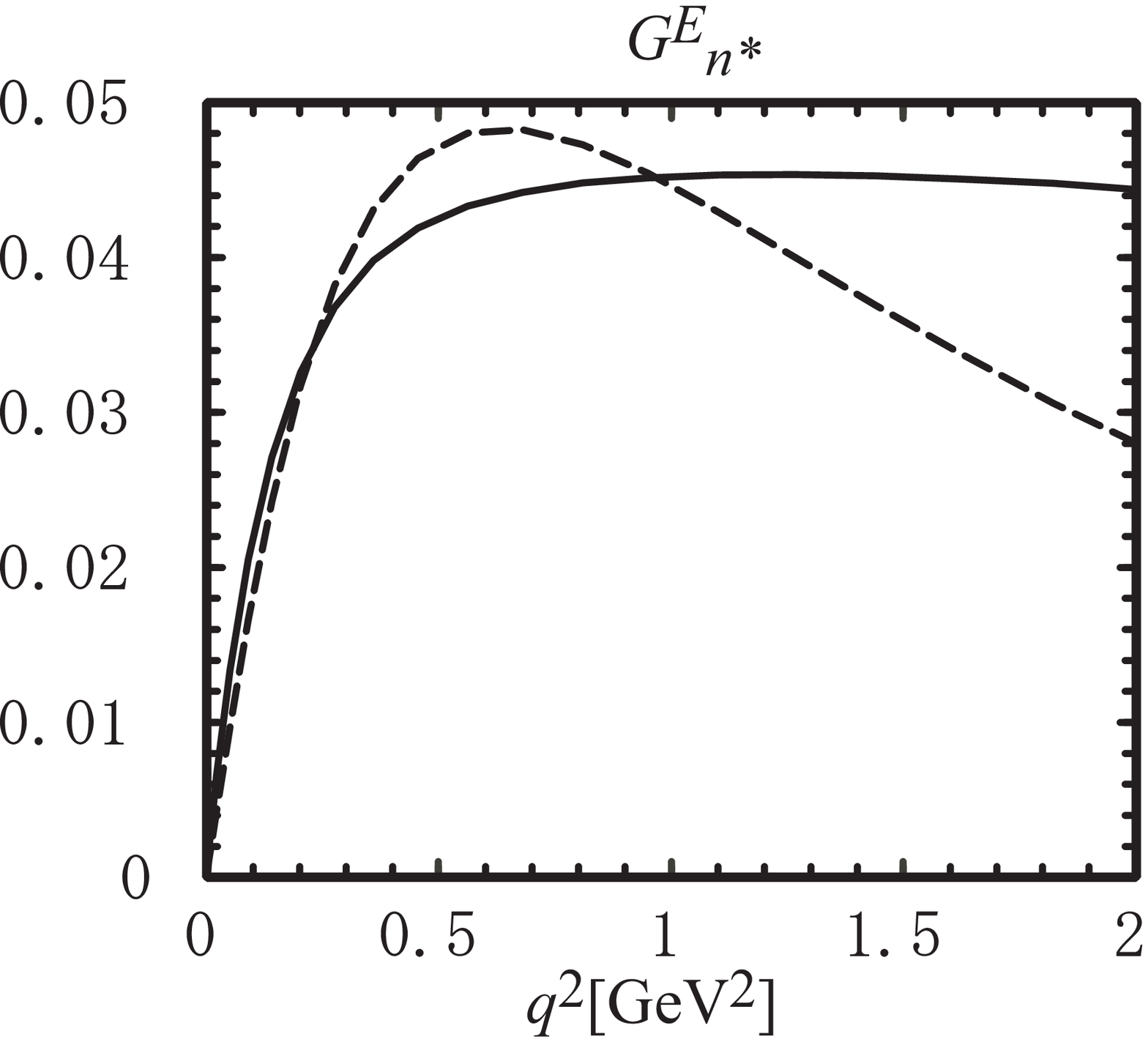}}
\begin{minipage}{11cm}
\caption{\small The electric form factors of the Roper resonance,  $p^*$ (left) and  $n^*$ (right). 
For a reference, the dashed lines are for the 
proton and neutron.}
\label{fig:ge_NR}
\end{minipage}
\end{figure}
Figures~\ref{fig:ge_NR} show the electric form factors of the Roper resonance. 
The slope of $p^*$ at $\vec{q}^{\;2}=0$ is comparable to that of the proton, 
but more precisely the slope of $p^*$ is slightly larger than that of $p$.
Therefore the charge radius of $p^*$ is larger than that of $p$. This is 
understood from that the charge radius of $p$ is dominated by the orbital 
motion of the quark in the scalar channel, while that of $p^*$ is dominated 
by the intrinsic size of the axial-vector diquark owing to the scalar 
(axial-vector) diquark dominance of the nucleon (Roper resonance). 
With respect to the neutron component, the charge radius of $n^*$ takes
almost same value as $n$. 
For the charge radius of $n$, the orbital motion of the $d$-quark with the charge 
$-1/3$ and the intrinsic size of the scalar diquark with $+1/3$ almost cancel.
The former is slightly larger than the latter,  therefore 
the charge radius of $n$ is negative.
Owing to the similar cancellation mechanism, the charge radius of $n^*$ is also negative.
In the case of $n^*$, the orbital motion of the quark and the intrinsic size of the
axial-vector diquark are almost same size. Hence the charge radius $n^*$ become 
positive when we employ the axial-vector diquark size larger than $0.8 $[fm].


In summary, we have investigated the electric form factors of the nucleon and 
Roper resonance in the chiral quark-diquark model.
We showed that the nucleon and Roper resonance have almost the same size in 
both the proton and neutron components.

In conventional pictures of the collective excitation of the Roper resonance,
the charge radii of the Roper resonance are larger than those 
of the nucleon both for the proton and neutron components. 
In the quark-diquark picture, the Roper resonance is a 
spin-partner of the nucleon with the different spin component.
In this case the charge radii of the Roper resonance become comparable 
to those of the nucleon and are smaller than those predicted in the collective pictures.

K.N is supported by National Science Council (NSC) of Republic 
of China under grants No. NSC96-2119-M-002-001.


\end{document}